\documentclass[aps,eqsecnum,showpacs]{revtex4}
\begin{document}

\preprint{{DOE/ER/40762-275}\cr{UMPP\#03-036}}

%\begin{flushright}
%DOE/ER/40762-276\\
%UMPP\#03-037
%\end{flushright}

\count255=\time\divide\count255 by 60
\xdef\hourmin{\number\count255}
  \multiply\count255 by-60\advance\count255 by\time
 \xdef\hourmin{\hourmin:\ifnum\count255<10 0\fi\the\count255}

\newcommand{\xbf}[1]{\mbox{\boldmath $ #1 $}}

\newcommand{\sixj}[6]{\mbox{$\left\{ \begin{array}{ccc} {#1} & {#2} &
{#3} \\ {#4} & {#5} & {#6} \end{array} \right\}$}}

\newcommand{\threej}[6]{\mbox{$\left( \begin{array}{ccc} {#1} & {#2} &
{#3} \\ {#4} & {#5} & {#6} \end{array} \right)$}}

\title{Excited Baryon Decay Widths in Large $N_c$ QCD}

\author{Thomas D. Cohen}
\email{cohen@physics.umd.edu}

\author{Daniel C. Dakin}
\email{dcdakin@physics.umd.edu}

\author{Abhinav Nellore}
\email{nellore@physics.umd.edu}

\affiliation{Department of Physics, University of Maryland,
College Park, MD 20742-4111}

\author{Richard F. Lebed}
\email{Richard.Lebed@asu.edu}

\affiliation{Department of Physics and Astronomy, Arizona State
University, Tempe, AZ 85287-1504}

%\date{\hourmin, \today}
\date{October, 2003}

\begin{abstract}
We study excited baryon decay widths in large $N_c$ QCD.  It was
suggested previously that some spin-flavor mixed-symmetric baryon
states have strong couplings of $O(N_c^{-1/2})$ to nucleons
[implying narrow widths of $O(1/N_c)$], as opposed to the generic
expectation based on Witten's counting rules of an $O(N_c^0)$
coupling.  The calculation obtaining these narrow widths was
performed in the context of a simple quark-shell model.  This
paper addresses the question of whether the existence of such
narrow states is a general property of large $N_c$ QCD.  We show
that a general large $N_c$ QCD analysis does not predict such
narrow states; rather they are a consequence of the extreme
simplicity of the quark model.
\end{abstract}

%11.15.Pg   Expansion for large numbers of components (e.g., 1/Nc expansion)
%12.39.-x   Phenomenological quark models
%13.75.Gx   Pion-baryon interactions
%14.20.Gk   Baryon resonances

\maketitle

\section{Introduction}

Large $N_c$ QCD~\cite{TH} has proved useful in predicting hadronic
properties.  This paper explores the important issue of excited
hadron lifetimes.  The behaviors of mesons and baryons appear to
differ radically at large $N_c$.  For example, Witten's well-known
large $N_c$ power counting rules specify that the decay widths of
mesons scale like $N_c^{-1}$, but the decay widths of baryons
above the ground-state band scale like $N_c^0$~\cite{EW}.  Mesons
are clearly narrow at large $N_c$; this helps explain why they are
visible in the spectrum.  However, generic baryons are not narrow
at large $N_c$. {\em Why can one still resolve excited baryons?}
Perhaps the answer is simply that the baryons actually dominating
the low-lying spectrum are exceptional states whose decay widths
are narrow in the large $N_c$ sense: They vanish in the large
$N_c$ limit.  In fact, it has been suggested that those baryons
transforming under the mixed-symmetric (MS) representation of the
spin-flavor group SU(2$N_f$) are narrow~\cite{PY}.  If this is
true, it is an exciting result; the observed narrow baryons are
generally assigned to such a representation in simple quark
models.  The purpose of the present paper is to investigate
whether the narrowness of such states is, in fact, a consequence
of large $N_c$ scaling in QCD.

We begin with a brief discussion of the relevant background.  A
seminal development in the study of large $N_c$ ground-state baryons
was the derivation of consistency conditions that constrained their
coupling to pions.  These consistency conditions were found by
examining pion-baryon scattering~\cite{DJM,GS,DM,J,DJM2,JL,Luty,CGO}.
A contracted SU($2N_f$) spin-flavor symmetry emerges from this
analysis, in which the ground-state band of baryons fills a completely
symmetric (S) representation ; it links various quantities such as
baryon axial-vector couplings, magnetic moments, and
masses~\cite{DJM,GS,DM,J,DJM2,JL,Luty,CGO}.  Recent interest has
focused on its application to the study of the $\Delta$ and the
$N$-$\Delta$ transition~\cite{BHL,BL1,BL2,JJM,Cohen,PY0}.

The success of the consistency condition method for describing
ground-state baryons naturally led to the question of whether excited
baryons could be understood in an analogous fashion.  Pirjol and Yan
(PY) developed just such an approach~\cite{PY}.  However, there is an
important distinction \cite{CL} between the treatments of ground-state
baryons and their excited cousins.  The scattering of pions off
ground-state baryons is physically realizable at large
$N_c$~\cite{Dwidth}.  On the other hand, the excited baryons are
resonances, so it is not immediately obvious how one can formulate a
pion-excited baryon scattering problem: The excited states decay and
cannot be used as targets in scattering experiments.  Thus, the
consistency condition approach is only applicable if there exist
excited baryons in the large $N_c$ world that are indeed narrow.  PY
tacitly assumed that such baryons exist.  As such, their general
model-independent predictions are not strictly valid.  However, if
there turns out to be a class of narrow excited baryons, then the
model-independent arguments can be applied to these in a legitimate
way.  Thus, the existence of a class of narrow baryon states at large
$N_c$ is also of importance in providing a theoretical justification
for the elegant model-independent analysis of Ref.~\cite{PY}, as
applied to at least some states.

Interestingly, Ref.~\cite{PY} itself provided an argument that there
exists a class of narrow excited baryons at large $N_c$.  It finds
that the baryons in the MS representation of the spin-flavor group
have widths of $O(N_c^{-1})$, which thus vanish at large $N_c$.  This
is in contrast to the generic large $N_c$ counting rule prediction, in
which the widths are $O(N_c^0)$~\cite{CL}.  Unfortunately, the
prediction of narrow decay widths for the MS states in Ref.~\cite{PY}
was {\em not\/} based directly on large $N_c$ consistency rules.
Rather, it arose from calculations done in the context of a simple,
nonrelativistic quark model.  The particular model employed was the
quark-shell model to be described in Section~\ref{sec:cons}.  Of
course, quark models have a long and distinguished history of
successful phenomenology.  In the large $N_c$ context, quark models
have been used to describe the lowest-lying excited baryon
states~\cite{Leb,Leb2,CGKM,G,SGS,CC,CC2}, and have revealed
interesting mass degeneracy patterns~\cite{PY,CL,PS}.  Recently, it
was shown that the quark-shell model is compatible with the more
realistic picture of excited baryons as resonances in meson-baryon
scattering, in the sense that both describe the same mass and width
degeneracy patterns~\cite{CL}.

The question of interest in this paper is whether the prediction
in Ref.~\cite{PY}, that there exist narrow excited baryon states,
is a direct consequence of large $N_c$ QCD.  It is useful to note
that a large number of the model-independent relations
found~\cite{DJM,GS,DM,J,DJM2,JL,Luty,CGO} were originally seen in
the context of soliton models \cite{AdkNap}, and indeed {\em
all\/} of these hold for quark models.  Thus, the issue is
whether the existence of narrow excited baryon states found in
the quark-shell model of Ref.~\cite{PY} similarly indicates a
general large $N_c$ result.  If so, this is an important general
result in understanding excited baryons.  In contrast, if the
prediction is merely an artifact of the particular choice of
model, then it is of far less import.  In this context, it is
important to note that the model used in Ref.~\cite{PY} is not
completely general.  It is limited in that it does not include
admixtures of different single-particle descriptions ({\it i.e.},
it neglects configuration mixing).

The issue of whether the MS excited baryons described in the simple
quark-shell model correspond to physical states in large $N_c$ QCD is
not addressed in Ref.~\cite{PY}.  If the states are physical, then a
previously unrecognized symmetry in large $N_c$ QCD is manifesting
itself and prevents the states from decaying rapidly.  In particular,
the narrowness depends upon a symmetry beyond the contracted SU($2
N_f$) spin-flavor symmetry deduced in
Refs.~\cite{DJM,GS,DM,J,DJM2,JL,Luty,CGO}.  In those works the
``spin'' in ``spin-flavor'' corresponds to the total angular momentum
of the baryon state in its rest frame.  As shown below, the narrow
states predicted in Ref.~\cite{PY} depend on {\em two\/} distinct
spin-flavor symmetries: one as before, in which the spin corresponds
to the total angular momentum of the baryon state in its rest frame,
and a second one in which the spin is purely associated with the spin
of the quarks.  While these two symmetries are identical for a
quark-shell model with all of the quarks in the orbital ground level
(s-wave), they differ for more general models.  The prediction of a
new symmetry emerging in the large $N_c$ limit for excited states is
certainly exciting; the issue, however, is whether it is real.

In this paper, we show that the seemingly narrow excited baryons are
in fact {\em not\/} a feature of large $N_c$ QCD; they are merely
artifacts of the simple quark model used.  Our paper is organized as
follows: We review the construction of baryon states in the
quark-shell model and the significance of key matrix elements in
Ref.~\cite{PY}.  Next, we use a quark-shell model Hamiltonian to show
that the states used in Ref.~\cite{PY} are not physical in large $N_c$
QCD.  In fact, the physical states are superpositions of both S and MS
representations of spin-flavor.

\section{Consistency conditions and the quark-shell model
\label{sec:cons}}

Pirjol and Yan's analysis~\cite{PY} of excited baryons proceeds
in a manner that is formally similar to that of Dashen, Jenkins,
and Manohar~\cite{DJM}.  Pions are scattered off excited baryons,
and the large $N_c$ counting rules are enforced for the total
scattering amplitude.  Any analysis of a conceivable scattering
event presupposes that the target is stable; in Ref.~\cite{PY},
the target was a p-wave baryon, and it was taken to be narrow at
large $N_c$.  This is a tenuous assumption unless the quark mass
is taken to be large, in which case the baryon has no possible
decay channels due to phase space limitations.  For the sake of
argument, we implicitly work in such a world in the following and
assume that results obtained in a such a world can be safely
extrapolated back to the physical world of light quarks.  As
noted in Section I, it is by no means clear that such a procedure
is legitimate, since the consistency conditions cannot really be
formulated for states that are unstable at large $N_c$.  However,
for the present purpose this procedure is adequate. States may be
narrow for one of two reasons: Either the phase space for decay
is small (or zero), or the meson-baryon coupling is small.  The
claim of Ref.~\cite{PY} is that MS states have a meson-baryon
coupling that goes as~$N_c^{-1/2}$, and thus even when the phase
space is of $O(N_c^0)$, it still produces narrow resonances.  If
the coupling turned out to be truly $O(N_c^{-1/2})$ as a result
of general large $N_c$ physics arguments, this counting would be
expected to hold regardless of whether the quark mass were taken
to be light or heavy enough to suppress the phase space for
decay.  Conversely, if one can show even in this world of heavy
quarks that the coupling to would-be decay channels (which are
now phase space suppressed) is generally $O(N_c^0)$ and not
$O(N_c^{-1/2})$ for {\em all\/} low-lying states in the spectrum,
then it is clear that large $N_c$ arguments by themselves do not
predict a class of weakly-coupled states.

The analysis of PY is of two parts: i) A set of consistency conditions
for couplings of excited baryons to mesons (analogous to the
consistency conditions for ground-state baryons) is derived.
Functional forms of relations that solve these conditions are proposed
and then verified.  ii) A simple nonrelativistic quark-shell model is
used to motivate the functional forms proposed for the
model-independent analysis.  As a result of this quark model analysis,
it is seen that the strength of the coupling between an excited baryon
and a meson plus a ground-state baryon depends on the symmetry class
of the excited baryon.

In the model-independent section, the aforementioned consistency
conditions were determined by imposing Witten's counting
rules~\cite{EW} on the following scattering processes at large
$N_c$:
\begin{eqnarray}
\pi^{a} + B(\mbox{{\rm s-wave}}) \rightarrow
\pi^{b}+ B'(\mbox{{\rm s-wave}}),\label{ss}\\
\pi^{a} + B(\mbox{{\rm p-wave}}) \rightarrow
\pi^{b}+ B'(\mbox{{\rm p-wave}}),\label{pp}\\
\pi^{a} + B(\mbox{{\rm p-wave}}) \rightarrow
\pi^{b} +B'(\mbox{{\rm s-wave}}),\label{ps}
\end{eqnarray}
where s-wave refers to the ground-state band of baryons modeled as
having all quarks in a spatial s-wave; p-wave refers to excited
levels that have quantum numbers consistent with a single quark
excited into a p-wave orbital.  The $a$ and $b$ are isospin
indices.

At each pion-baryon vertex, the baryon axial-vector current is
derivatively coupled to the pion field:
\begin{equation}
\langle B'|\bar q \gamma^i \gamma_5 \tau^a q|B\rangle \, \partial^i
\pi^a/f_{\pi}=N_c\langle B'|X^{ia}|B\rangle \, \partial^i
\pi^a/f_{\pi},\label{X}
\end{equation}
where the current matrix element is parameterized in terms of an
irreducible tensor operator $X^{ia}$ and an explicit power of $N_c$.
An $O(N_c^0)$ coupling $g(X)$ usually included on the right-hand side
of this equation is absorbed into the definition of $X$ for
convenience.  The scattering amplitude for the two leading-order
tree-level diagrams for Eq.~(\ref{ss}) is ${\cal A}\sim
(N_c/f_{\pi})^2[X^{jb\dagger}, X^{ia}]$ and thus is naively $O(N_c)$
since $f_{\pi}\sim O(N_c^{1/2})$.  Such scaling contradicts the Witten
$N_c$ power-counting prediction (as well as unitarity at large $N_c$),
which require it to be $O(N^0_c)$; one is led to the conclusion that
$[X^{jb\dagger}, X^{ia}]=0$ in the large $N_c$ limit.  The vanishing
of the commutator is the leading-order consistency condition for
ground-state baryons and is the key to the contracted SU(4)
algebra~\cite{GS,DJM}.

This procedure can be extended to the process in Eq.~(\ref{pp}) if the
current matrix element is parameterized in terms of a new operator
$Z^{ia}$: $\langle B'|\bar q \gamma^i \gamma_5
\tau^a q|B\rangle=N_c\langle B'|Z^{ia}|B\rangle$.  As in the above
case, the scattering amplitude apparently diverges at large $N_c$ in
the absence of cancellations, and thus consistency requires that
\begin{equation}
[Z^{jb\dagger}, Z^{ia}]=0\label{conZ}
\end{equation}
in the large $N_c$ limit.  This condition is analogous to that for
$X^{ia}$, meaning that solutions for $Z^{ia}$ also fill irreducible
representations of a contracted SU(4) algebra.  These representations
can be labeled by the magnitude of a spin vector $\vec{\Delta}$ such
that $\vec{\Delta}=\vec{I}+\vec{J}$ (but only in the sense that
allowed eigenvalues of $\vec{\Delta}$ are determined by the vector
addition rule; indeed, Ref.~\cite{PY} uses a relative minus sign in
this definition).  Note that this operator (denoted $\vec{K}$ in
Ref.~\cite{CL}) has a very simple interpretation in terms of chiral
soliton models, in which case $\vec{\Delta}=\vec{I}+\vec{J}$ in a true
vector operator sense, as one has from studies of a canonical hedgehog
configuration, where the combined operator is called the ``grand
spin.''

To extend this procedure to the process in Eq.~(\ref{ps}), one must
introduce two new operators, $Y^a$ and $Q^{ij,\,a}$, in order to
parameterize the current matrix elements between an s-wave and p-wave
baryon:
\begin{eqnarray}
\langle B'|\bar q \gamma^0 \gamma_5 \tau^a
q|B\rangle & = & N_c^{1/2}\langle B'|Y^{a}|B\rangle , \label{Y}\\
\langle B'|\bar q \gamma^i \gamma_5 \tau^a q|B\rangle & = &
N_c^{1/2}q^j\langle B'|Q^{ij,\,a}|B\rangle, \label{Q}
\end{eqnarray}
where $q^{\mu}$ is the momentum of the current and $|B'\rangle$
indicates the ground-state baryon.  These expressions differ from
those in Ref.~\cite{PY} by the absorption of possible additional
$N_c$ powers and coefficients $g(Y,Q)$ into the right-hand sides,
which can be accommodated by explicit rescaling of $Y$ and $Q$.
The scattering amplitude for Eq.~(\ref{ps}) still violates the
Witten power-counting prediction if the $Y$ and $Q$ matrix
elements scale as $N_c^{-1/2}$ or larger (note that generic
Witten counting rules suggest that $Y$ and $Q$ scale as $N_c^0$),
and in these cases consistency requires that
\begin{eqnarray}
X^{ia}Y^{b \dagger} - Y^{b \dagger} Z^{ia}=0
& \hspace{.1in}\textrm{and}\hspace{.1in} &
X^{ib \dagger}Y^a - Y^aZ^{ib \dagger}=0 ,\label{conY}\\
X^{ia}Q^{jk,\,b\dagger} - Q^{jk,\,b\dagger}Z^{ia}=0
& \hspace{.1in}\textrm{and}\hspace{.1in} &
X^{kb \dagger}Q^{ij,\,a} - Q^{ij,\,a}Z^{kb\dagger}=0\label{conQ}.
\end{eqnarray}
The set of consistency conditions in Eqs.~(\ref{conZ}), (\ref{conY}),
(\ref{conQ}) form the basis of Pirjol and Yan's model-independent
analysis.  Matrix elements of the operators $Z$, $Y$, and $Q$ between
baryon states $|J, J_3, I, I_3, \Delta\rangle$ can be found by solving
this set.  We repeat for emphasis that the minimum set of quantum
numbers needed to construct solutions to the consistency conditions is
$\{J, J_3, I, I_3, \Delta\}$.  The $Y^a$ matrix element is, for
example,
%
%\begin{eqnarray}
%\lefteqn{\langle J' \! = \! I', J_3', I_3'|Y^a|J, J_3, I, I_3,
%\Delta\rangle =}\nonumber\\
%&& g_Y (-1)^{J-I}\sqrt{2I+1} \, \langle I'I_3'|I,1;I_3,a\rangle
%\, \delta_{JJ'} \delta_{J_3J_3'}\left\{
%\begin{array}{ccc} I' & 1 & I\\ \Delta & J & 0 \end{array}
%\right\},\label{solnY}
%\end{eqnarray}
%
\begin{equation}
\langle J' \! = \! I', J_3', I_3'|Y^a|J, J_3, I, I_3,
\Delta\rangle = g_Y (-1)^{1+2J} \sqrt{\frac{2I+1}{3(2J+1)}} \,
\langle I'I_3'|I,1;I_3,a\rangle \, \delta_{JJ'} \delta_{J_3J_3'}
\delta_{\Delta 1}, \label{solnY}
\end{equation}
where the fact that $|B' \rangle$ is a ground-state baryon
imposes the condition $I^\prime \! = \! J^\prime$.  The constant
$g_Y$ encodes the overall strength, including any overall
nontrivial $N_c$ scaling.  If the system scales according to the
generic Witten rule, then $g_Y \! \sim \! N_c^0$.  If for some
special class of states the coupling is characteristically
smaller, then $g_Y$ is smaller than $O(N_c^0)$.
Reference~\cite{PY} calculates the matrix elements in the
quark-shell model and finds the same spin-flavor structure as
given by Eq.~(\ref{solnY}), regardless of the symmetry of the
state $|B \rangle$.  However, the $N_c$ dependence of the
coefficient $g_Y$ in the quark-shell model is found to depend
upon the symmetry of the excited states, with $g_Y \! \sim \!
N_c^0$ for spin-flavor S states and $g_Y \! \sim \! N_c^{-1/2}$
for MS states.  It is this scaling for the MS states that leads
to the prediction of narrow baryon resonances.

In the simple quark-shell model of Ref.~\cite{PY}, each excited baryon
is treated as a single orbitally excited quark with angular momentum
$\ell$ acting on top of a spin-flavor symmetric core of $N_c - \!  1$
quarks.  The Pauli principle requires that the complete wave function
describing the baryon is antisymmetric under the exchange of any two
quarks.  The singlet color wave function is fully antisymmetric;
accordingly, the space and spin-flavor wave functions together must be
symmetric.  Only symmetric or mixed-symmetric spatial wave functions
can be constructed when a single quark is excited.  Therefore, the
spin-flavor wave functions of the quark-shell model states are either
symmetric or mixed-symmetric under exchange.  It is worth noting that
in the present context ``spin-flavor'' refers to the spin and flavor
of the quarks and not of the baryons.

We focus on the nonstrange states of SU(4), in which the distinction
between the MS and the S representations is clean: Spin and isospin
are related by $S \! = \! I$ for the S case and $|S-I|\leq 1$ for the
MS case.  This distinction may be neatly encoded by introducing the
the concept of $P$-spin~\cite{PY}: with $P \! = \! 0$ for the S case
and $P \! = \! 1$ for the MS case.  Thus, treating $\vec{P}$ as though
it were a true angular momentum, one sees that the single triangle
rule $\delta (SIP)$ characterizes both permutational symmetries.

The matrix elements of $Z$, $Y$, and $Q$ in the quark model are
obtained by defining the currents and constructing the baryon states
in quark model language.  Up to overall multiplicative constants of
order unity, the currents in the quark model are:
\begin{eqnarray}
N_cZ^{ia} & = & \sigma^{i} \otimes \tau^{a},\label{Z}\\
N_c^{1/2}Y^{a}  & = &  \frac{1}{\sqrt{3}}\sum_{j=-1}^{+1}(-)^{1-j}
\sigma^{j}r^{-j}\otimes\tau^{a},\\
N_c^{1/2}Q^{ka} & = & \sum_{i,j=-1}^{+1}\langle2,k|1,1;j,i\rangle
\, \sigma^{j}r^{i}\otimes\tau^{a}.
\end{eqnarray}
The $\sigma$, $\tau$, and $r$ operators act on the quark's spin,
isospin, and orbital degrees of freedom, respectively.

Reference \cite{PY} constructs quark model states of good (quark) spin
($S$, $m_S$), isospin ($I$, $I_3$), and orbital angular momentum
($\ell$, $m_\ell$) in such a manner that each is either symmetric or
mixed-symmetric under spin-flavor.  It is not clear at the outset how
such states should be interpreted.  Either these states may be taken
to be eigenstates of some unspecified Hamiltonian ${\cal H}$ that is
assumed to model QCD, or they may be taken to be merely a convenient
basis that allows one to enumerate the possible physical states.  If
they are eigenstates of ${\cal H}$, then the quantum numbers
specifying the states must be associated with operators that commute
with ${\cal H}$.  If this is the case, then the eigenstates of ${\cal
H}$ include the narrow excited states predicted in Ref.~\cite{PY} and
incorporate a new large $N_c$ symmetry that makes these states stable.
This ostensible symmetry arises because ${\cal H}$ commutes separately
with the (quark) spin and with the total angular momentum.

Note that if the states are merely used to form a basis, then one is
faced with the issue of determining the scaling of mixing between the
basis states.  If each physical state is predominantly a single basis
state (with admixtures of other states characteristically suppressed
in the large $N_c$ limit), the system then acts much as it would for
the case where the states are treated as eigenstates of a Hamiltonian
that mocks up QCD: The physical eigenstates that are predominantly
mixed-symmetric are then narrow.  In contrast, if the mixing is
$O(N_c^0)$, then the concept of a state that is predominantly
mixed-symmetric is ill-defined, and all states allowed by phase space
have widths that go as $N_c^0$.

It should also be observed that the quantum numbers $\{P, S, m_S, I,
I_3, \ell, m_\ell\}$ denoting these states are different from those
used in the (model-independent) consistency condition method, $\{J,
J_3, I, I_3, \Delta\}$.  In particular, it should be noted that the
there is no analog for the $P$ quantum number (which specifies the
nature of the spin-flavor symmetry of the quark model state) in the
model-independent analysis.  This raises the obvious question of
whether or not the concept of the $P$-spin has a well-defined meaning
at large $N_c$ outside the context of the simple quark model.

The matrix elements of the above currents can be calculated with the
quark model states described above.  A lengthy calculation in
Ref.~\cite{PY} revealed the $N_c$ scaling of the matrix elements of
$Y$ (and $Q$) as well as the detailed spin-flavor structure; it was
found that
\begin{equation}
\langle P' \! = 0, \ell' \! = 0\left|Y, Q \right|P \! =0,
\ell\neq 0\rangle \sim N_c^0 ,
%\hspace{.1in}\textrm{and}
%\hspace{.1in}\langle P' \! = 0,
%\ell' \! = 0\left|Q\right|P \! = 0,
%\ell\neq 0\rangle \sim N_c^{1/2},\nonumber\\
\end{equation}
while
\begin{equation}
\langle P' \! = 0,\ell' \! = 0\left|Y,Q \right|P \! = 1, \ell\neq
0\rangle \sim N_c^{-1/2} ,
%\hspace{.1in}\textrm{and}
%\hspace{.1in}\langle P' \! = 0,
%\ell' \! = 0\left|Q\right|P \! = 1, \ell\neq 0\rangle
%\sim N_c^{0}\,,
\label{MStoS}
\end{equation}
where the states are labeled by their $P$-spin and orbital angular
momentum.

The excited baryon decay widths are determined by squaring the matrix
elements of the physical states and dividing by the pion decay
constant $f_\pi^2$ and including the appropriate phase space factor.
Using that $f_\pi^{2} \! \sim \! O (N_c)$ and the phase space is
$O(N_c^0)$, it is straightforward to determine the scaling of the
decay widths.  If one assumes that quark model assignments of states
correspond to physical states (with only small admixtures of states of
different quark model symmetries), then the decay width of an MS
baryon ($P \! = \! 1$) is $O(N_c^{-1})$, while the decay width of an S
baryon ($P \! = \! 0$) is of order $N_c^{0}$.  In the large $N_c$
limit, the former vanishes.  This result is of import; it implies that
\textit{MS excited baryon states are narrow}.  This would neatly
explain the phenomenological fact that certain baryons, like mesons,
are narrow enough to discern and would render the consistency argument
of Ref.~\cite{PY} valid.  However, these desirable results depend on
the physical baryon states corresponding to the quark model states in
terms of their quantum numbers.  Thus, one must face the question of
whether they do.

\section{Spin-flavor symmetry breaking and baryon widths \label{mix}}

In this section we focus on the issue of whether the physical states
truly correspond to the simple quark model states of Ref.~\cite{PY},
which in effect is the question of whether narrow excited baryons are
realized in large $N_c$ QCD.  If this result is generic to large $N_c$
QCD, one would expect it to be seen in all models that correctly
encode large $N_c$ physics.  Thus, to disprove it we need only find
some model that encodes the correct large $N_c$ scaling rules for
which it is untrue.  Here we consider a fairly general quark-shell
model Hamiltonian that shares some essential properties with the QCD
Hamiltonian.  In particular, we consider the most general quark model
for which the number of quarks in a given orbital is well defined.  In
practice, this restriction means one excludes operators that remove
quarks from one orbital and place them in different orbitals.  We
impose this restriction to keep the model tractable.  Note, however,
that even with this restriction, this Hamiltonian is considerably more
general then the Hamiltonian implicitly used to construct the states
in the previous section.

Large $N_c$ scaling rules {\em greatly\/} restrict the number of
possible operators that contribute at $O(N_c^0)$ or larger (and hence
that can contribute in the large $N_c$ limit~\cite{Leb,Leb2}).  For
example, as shown in Ref.~\cite{Leb,Leb2}, the {\em only\/} operators
that contribute at $O(N_c^0)$ for states with a single excited quark
are given by:
\begin{equation}
{\cal H} = c_1 \openone + c_2 \ell \! \cdot \! s + c_3 \ell^{(2)}g
G_c/N_c ,
\end{equation}
where the $\ell$, $s$, $\ell^{(2)}$, and $g$ are the orbital, spin,
$\Delta \ell=2$ tensor, and combined spin-flavor (Gamow-Teller)
operators, respectively, acting only on the excited quark, while $G_c$
is the combined spin-flavor operator acting only on the core of
unexcited quarks.  The coefficients have the following scaling rules:
\begin{equation}
c_1 \sim N_c^1 ,\;\;\; c_2 \sim N_c^0 , \;\;\; c_3 \sim N_c^0 \;.
\label{cscale}
\end{equation}
The scaling of $c_1$ is a bit subtle.  Most of the contribution to
$c_1$ comes from the unexcited quarks in the core.  Thus $c_1 = M_N +
\delta c_1$ with $\delta c_1 \sim N_c^0$.  In general, each
coefficient contains corrections at all subleading powers of $N_c$.

Consider, as an example, the operator associated with $c_2$.  If this
operator induces significant [$O(N_c^0)$] mixing between the S and MS
states of the basis described above, then this model---which correctly
encodes the large $N_c$ scaling rules---does not automatically give
excited baryons that are weakly coupled at large $N_c$.  To begin,
note that the $\ell \! \cdot \! s$ term does not commute with the spin
operator $\textbf{S}$; {\it i.e.}, $m_S$ is not generally a good
quantum number for the Hamiltonian eigenstates.  Thus, the operator
does induce mixing between the states enumerated above.  The central
question becomes the scale of this mixing.

Suppose one considers only states for which the excited quark is in an
orbital with $\ell \! \ne \! 0$ (The case of $\ell \! = \! 0$ is
special and is discussed below).  This implies that the $\ell
\! \cdot \! s$ operator mixes states of different spin-flavor
symmetry.  Consider a state labeled by total angular momentum
($J$,$J_3$), total isospin ($I$, $I_3$), total (quark) spin ($S$), and
$P$-spin: $|JJ_3;II_3(\ell, S=I+\rho)[P]\rangle$.  The $\rho$
(introduced in Refs.~\cite{Leb,Leb2}) plays a role similar to the
$P$-spin of Sec.~\ref{sec:cons}.  It is a
\textit{number\/} that equals either $\pm1$ or 0 for the mixed-symmetric
case ($P=1$), or 0 for the symmetric case ($P=0$).  The states so
labeled are identical to the ones enumerated in Sec.~\ref{sec:cons}.
The matrix element of $\ell \! \cdot \! s$ that connects two states in
this basis of equal $JJ_3$, $II_3$, and $\ell$ but different $P$-spin
({\it i.e.}, symmetry) is written as $\langle \ell \! \cdot
\!  s\rangle_{\rho}\equiv\langle JJ_3;II_3(\ell,
S^\prime \! = \! I+\rho^\prime)[1]\left|\ell \! \cdot \!
s\right|JJ_3;II_3(\ell, S \! = \! I)[0]\rangle$.  We follow the
methods of Ref.~\cite{Leb}, but note a more concise expression than
their Eq.~(A7):
\begin{equation}
\langle \ell \! \cdot \! s\rangle = (-1)^{J-I+\ell}
\sqrt{\frac{3}{2}}\sqrt{\ell (\ell+1) (2\ell+1) (2S+1) (2S^\prime+1)}
\left\{ \begin{array}{ccc} \ell & \ell & 1 \\ S & S^\prime & J
\end{array} \right\} \sum_{\eta = \pm 1} c_{\rho \eta} c_{\rho^\prime
\eta} (-1)^{(1-\eta)/2} \left\{ \begin{array}{ccc} 1 & \frac 1 2 &
\frac 1 2 \\ I \! + \! \frac \eta 2 & S^\prime & S \end{array} \right\} .
\end{equation}
From here, we set $\rho = 0$ (meaning that $S \! = \! I$ for the S
state in the ket) and calculate this matrix element for each value of
$\rho^\prime$ in the mixed-symmetric bra.  For $\rho^\prime=0$, we
have:
\begin{equation}
\langle \ell \! \cdot \! s\rangle_0=
(-1)^{J-I+\ell+1} \sqrt{\frac{3}{2}}
\sqrt{\ell (\ell+1) (2\ell+1)} (2S+1)
c^{\texttt{MS}}_{0-}c^{\texttt{MS}}_{0+}
\left\{ \begin{array}{ccc} \ell & \ell & 1 \\ S & S & J \end{array}
\right\} \left[ \left\{ \begin{array}{ccc}
1 & \frac 1 2 & \frac 1 2 \\ I \! + \! \frac 1 2 & S & S
\end{array} \right\} + \left\{ \begin{array}{ccc} 1 & \frac 1 2 &
\frac 1 2 \\ I \! - \! \frac 1 2 & S & S \end{array} \right\} \right] .
\label{zero}
\end{equation}
%
%\begin{eqnarray}
%\langle \ell \! \cdot \! s\rangle_0=
%\frac{2S+1}{2}\sum_{\ell_s}^{\ell\pm\frac{1}{2}} \left[
%\ell_s(\ell_s+1)-\ell(\ell+1)-\frac{3}{4} \right]
%(2\ell_s+1)c^{\texttt{MS}}_{0-}c^{\texttt{MS}}_{0+}\left[ \left\{
%\begin{array}{ccc} I-\frac{1}{2} & \frac{1}{2} & S\\ \ell & J & \ell_s
%\end{array} \right\}^2 - \left\{
%\begin{array}{ccc} I+\frac{1}{2} & \frac{1}{2} & S \\ \ell & J &
%\ell_s\end{array}\right\}^2\right]. \label{zero}
%\end{eqnarray}
%
The coefficients $c^{\texttt{MS}}_{0-}$ and $c^{\texttt{MS}}_{0+}$
are given by
\begin{eqnarray}
c^{\texttt{MS}}_{0-}&=&-\sqrt{\frac{(S+1)(N_c-2S)}{N_c(2S+1)}} \; ,
\nonumber\\
c^{\texttt{MS}}_{0+}&=&+\sqrt{\frac{S[N_c+2(S+1)]}{N_c(2S+1)}}\;.
\end{eqnarray}
For $\rho^\prime=\pm1$, we have:
\begin{equation}
\langle \ell \! \cdot \! s\rangle_{\pm 1} =
(-1)^{J-I+\ell+1} \sqrt{\frac{3}{2}}
\sqrt{\ell (\ell+1) (2\ell + 1) (2S \! + \! 1) (2S \! + \! 1 \pm 2)}
\, c^{\texttt{MS}}_{0\mp}
\left\{ \begin{array}{ccc} \ell & \ell & 1 \\ S & S \! \pm \! 1 & J
\end{array} \right\}
\left\{ \begin{array}{ccc} 1 & \frac 1 2 & \frac 1 2 \\
I \! \pm \! \frac 1 2 & S \! \pm \! 1 & S \end{array} \right\} .
\label{pm}
\end{equation}
%
%\begin{eqnarray}
%\langle \ell \! \cdot \!
%s\rangle_{\pm1}=(\mp)\frac{\sqrt{(2S+1)(2S+1\pm2)}}{2}\nonumber\hspace{4in}\\
%\times\sum_{\ell_s}^{\ell\pm\frac{1}{2}} \left[
%\ell_s(\ell_s+1)-\ell(\ell+1)-\frac{3}{4} \right]
%(2\ell_s+1)c^{\texttt{MS}}_{0\mp}\left\{
%\begin{array}{ccc} I\pm\frac{1}{2} & \frac{1}{2} & S\pm1\\ \ell & J & \ell_s
%\end{array} \right\}\left\{
%\begin{array}{ccc} I\pm\frac{1}{2} & \frac{1}{2} & S \\ \ell & J &
%\ell_s\end{array}\right\}. \label{pm}
%\end{eqnarray}

In general, evaluation of these formulas for appropriate values of
$S$, $I$, $J$, and $\ell$ yields nonzero values.  The nonvanishing of
MS $\to$ MS matrix elements is confirmed in Refs.~\cite{Leb,Leb2}.
Moreover, in all of these cases the matrix element of $\langle \ell \!
\cdot \! s\rangle$ is $O(N_c^0)$.  Combining this with the scaling of
Eq.~(\ref{cscale}) for the strength of the operator, implies that the
$\ell \! \cdot \!  s$ term in the Hamiltonian connects states of
different spin-flavor symmetry with strength of $O(N_c^0)$.  Due to
this mixing, an energy eigenstate of the quark-shell model cannot be
described as having a well-defined spin-flavor symmetry.  Thus, for
this class of models there is no special set of weakly-coupled excited
baryons at large $N_c$ (at least for $\ell \! \ne \! 0 $).  Since
these models encode generic large $N_c$ scaling, one concludes that
generic large $N_c$ scaling rules by themselves do not imply that a
set of weakly coupled states exists.  Returning now to a world where
the quarks are light enough that the decays are permitted by phase
space [which is $O(N_c^0)$], one concludes that there is no generic
argument why such states should be narrow.

Note that an analogous argument could be made using the $\ell^{(2)}g
G_c/N_c$ term in the Hamiltonian, which also leads to mixing of
$O(N_c^0)$.  One might wonder whether there is some way to evade this
conclusion by having some type of cancellation between the
$\ell^{(2)}g G_c/N_c$ and $\ell \! \cdot \! s$ terms.  However, the
two terms generically have nothing to do with each other; their ratio
is not fixed by large $N_c$ arguments.  Moreover, although the
operators commute at leading order, they are distinct---their matrix
elements are not proportional to each other, even at leading
order~\cite{CL,Leb,Leb2,PS}.  Hence, the only way for them to cancel
generally is if they are both zero.

The conclusion that S and MS configurations are mixed in models of
this type is valid for all cases except where the excited quark is in
an $\ell \! = \! 0$ orbital.  However, it is clear that all of the
matrix elements of $\ell \! \cdot \! s$ ($\Delta \ell \! = \! 1$) and
$\ell^{(2)}g G_c/N_c$ ($\Delta \ell \! = \! 2$) between states with
$\ell \! = \! 0$ are zero.  Thus, the argument presented above does
not exclude the possibility of weakly coupled, and hence narrow, $\ell
\!  = \! 0$ MS excited states.  We note, however, that the quark model
considered in this paper, though more general than that implicitly
used to construct the basis states, is by no means the most general
one that one can consider.  In particular, one can consider models
with configuration mixing---that is, in which the physical states are
admixtures of different single-particle descriptions~\cite{Ripka}.
Such operators can induce admixtures between the S and MS states at
$O(N_c^0)$.  It is easy to see how this can come about.  An allowable
operator can mix a state with a quark in an excited $\ell \! = \! 0$
orbital and a state with a quark in an $\ell
\! = \! 1$ orbital that has total angular momentum (spin plus orbital)
equal to 1/2.  Such mixing violates no symmetries of the system and is
allowable at $O(N_c^0)$.  Once such a state admixes with the $\ell
\! = \! 1$ orbitals, the previously considered operators induce mixing
between the S and MS spin-flavor components.

The case of $\ell \! = \! 1$ presents its own subtlety, the well-known
problem in many-body physics of spurious modes associated with broken
symmetries~\cite{Ripka}.  In fact, the existence of spurious modes
does not alter the conclusions drawn above for the $\ell \! = \! 1$
case, as discussed in Appendix~\ref{app}.

\section{Conclusion}

In this paper we have explored the issue of whether excited baryons
with mixed-symmetric spin-flavor wave functions have decay widths that
vanish in the large $N_c$ limit.  As noted earlier, if previous claims
that a set of states are automatically narrow at large $N_c$ were in
fact generic, it would be of real significance both theoretically and
phenomenologically.  On the other hand, the assertion that such states
are narrow is based on calculations with a very simple quark model.
In this work we showed that the purported narrowness of these states
is an artifact of the simple quark model used in the calculations and
not a generic feature of large $N_c$ QCD.  This was shown in the
context of a slightly more general class of quark models that encode
generic large $N_c$ scaling rules by a demonstration that excited
baryons cannot be assigned a well-defined, fixed spin-flavor symmetry;
the symmetry configurations are admixed at $O(N_c^0)$.  This implies
that the relative narrowness of baryon states observed in nature
cannot be simply attributed to large $N_c$ scaling behavior.  It also
implies that the general model-independent analysis of Ref.~\cite{PY}
is not strictly correct: Without a scattering target of narrow states,
the large $N_c$ consistency condition analysis is not applicable.
Fortunately, many of the conclusions of this analysis remain correct
despite this, such as the predicted pattern of degeneracies~\cite{CL}.

\section*{Acknowledgments}
T.D.C.\ and R.F.L.\ wish to thank Paolo Bedaque and the ``Effective
Summer at Berkeley'' program for their hospitality; the research
atmosphere at this program greatly facilitated the collaboration that
resulted in this paper.  The work of T.D.C., D.C.D., and A.N.\ was
supported in part by the U.S.\ Department of Energy under Grant No.\
DE-FG02-93ER-40762.  The work of R.F.L.\ was supported by the National
Science Foundation under Grant No.\ PHY-0140362.

\appendix
\section{The role of spurious modes \label{app}}

The quark-shell model possesses degrees of freedom associated with the
motion of all of the particles.  The total motion may always be
separated into center-of-mass (c.m.) motion and internal motion.
Moreover, since the underlying dynamics is translationally invariant,
the c.m.\ motion may always be quantized in states of good total
momentum.  Thus, in principle for a quark model with $N_c$ quarks, the
internal dynamics is only associated with the $N_c- \! 1$
displacements from c.m.  Since the model for the excitation spectrum
as written includes the positions of all $N_c$ quarks as explicit
degrees of freedom, there is in principle redundant information.  This
raises the obvious question of how one can separate the motion of the
c.m., which is spurious from the point of view of internal dynamics,
from the internal dynamics of interest.  Since the spurious c.m.\
motion is vectorial in nature---it is associated with the total
momentum $\vec{P}$---one expects that it manifests itself only in
channels that transform vectorially, which means the $\ell \!  = \! 1$
channels.

The issue of how the spurious motion is dealt with depends in part on
how the model under consideration is derived.  One may view the
quark-shell model used in this paper as being obtained from some
underlying translationally-invariant quark model with $N_c$ quarks
undergoing mutual interactions.  Traditionally, one approximates this
model with some type of self-consistent single-particle potential
model state such as the Hartree-Fock state~\cite{Ripka}.  The idea is
to choose a single-particle description that is optimal in the sense
of capturing the maximum amount of the underlying physics.  One then
attempts to include systematically the physics excluded by the
single-particle description.  Note that this single-particle
description necessarily breaks translational invariance, since one
cannot have a single-particle description with nontrivial internal
dynamics that simultaneously is an eigenstate of the c.m.\ momentum
(since the remaining degrees of freedom in this approach remain
inert).  From the perspective of such a model state, the spurious
c.m.\ motion is now associated with the fact that the approximation
scheme breaks translational symmetry.

It is obvious that if one treats the internal dynamics exactly, then
any ``excitation'' of the spurious c.m.\ motion on top of some
internal state does not alter the internal dynamics and necessarily
leads to a state whose internal energy is degenerate with the original
state.  There are certain approximation schemes that automatically
give zero-energy excitations for motion associated with the spurious
c.m.\ motion~\cite{Ripka}.  Such approximations are referred to as
``conserving approximations'' if they provide such an order-by-order
decoupling in the approximation scheme.  An example of such an
approximation is an RPA treatment above a Hartree-Fock trial
state~\cite{Ripka}.

Unfortunately, the quark-shell model does not correspond to a
truncation of a conserving approximation of an underlying
translationally invariant model at some consistent order.  Rather, it
is the form one obtains via the truncation of a Tamm-Dancoff type
expansion; such expansions are not conserving in an order-by-order
sense.  A true separation of the spurious motion from the internal
motion occurs if one includes all possible spatial orbitals (including
the continuum) and all possible N-body forces (and if the coefficients
of all of these are obtained in a consistent way from the underlying
translationally-invariant system).  This is not done for practical
reasons; the system so obtained is not computationally tractable.  The
effect of the {\it ad hoc\/} truncation is the contamination of the
physically interesting physics with unphysical spurious motion.  To
the extent that the truncation is not too severe, such contamination
is modest in that the physically interesting mode only has small
admixtures of the spurious motion, and the predicted physical motion
is accurate to good approximation.

There is one obvious drawback to such a scheme, apart from the
necessary numerical inaccuracy induced by truncation.  The Hilbert
space contains both physical and spurious motion, and the final
answers contain mixtures of both.  One hopes that one set of modes is
mostly physical (and can be identified with the physical result),
while the other set is mostly spurious and can be discarded.  One
therefore needs some method to discern which set is which.  The
obvious approach is to perform the calculation and {\it a
posteriori\/} identify the modes that are largely spurious.  One
natural way to identify them is to pick out anomalously low-lying
modes, recalling that true spurious modes correspond to zero-energy
excitations.

The problem of spurious modes as discussed here is generic in
quark-shell models, and in principle is totally divorced from the
issue of relevance to this paper, that of whether the S and MS states
mix strongly in forming the physical states.  One should follow the
strategy given above: Using the model, calculate all of the $\ell \!
= \! 1$ modes and discard the ones that are mostly spurious.  The
general arguments given in Sec.~\ref{mix} show that the states so
generated have strong mixing, and this is sufficient for our purpose.
One expects the physical $\ell \! = \! 1$ state to exhibit strong
mixing between states of different spin-flavor symmetries, and hence
one expects the states to have widths of $O(N_c^0)$.

However, there is one exceptional situation in which these two issues
appear to be related.  Consider a simple single-particle quark model
with orbitals given by harmonic oscillator states.  Since this model
has no spin-dependent interactions, the quark spin is a good quantum
number, and states may be labeled according to spin-flavor.  It is a
straightforward exercise to show that the spin-flavor S state
associated with a single quark in the lowest p-wave orbital is
entirely associated with spurious c.m.\ motion~\cite{Close}.  The
underlying mathematical reason is simply that the lowest harmonic
oscillator p-wave state is proportional to $\vec{r}$ times the ground
state.  Based on this experience, one might be under the impression
that the spurious mode is generally the lowest p-wave S state.  This
is not the case.  In fact, even for the simple case with harmonic
oscillator orbitals it is clear that separation into spurious and
physical modes is not dynamically correct.  The spurious $\ell \! = \!
1$ modes in such a model are degenerate with the physical $\ell \! =
\! 1$ modes rather than, as one might expect, with the ground state.
Indeed, the spurious modes in this model correspond to the c.m.\
oscillating harmonically with the excitation frequency.  Physically,
of course, the c.m.\ moves with a constant velocity.

Thus, we conclude that our previous arguments are valid even for $\ell
\! = \! 1$ states.  We note, however, that the entire
issue is moot.  In the end, the problem of spurious states is a
disease of quark-shell models.  They appear as
translationally-invariant quark models with {\it ad hoc\/}
truncations, and the disease is associated with the truncations.  Of
course, we introduced these models following the treatment of
Ref.~\cite{PY} and then generalized the models to show that even
quark-shell models have $O(N_c^0)$ mixing between different
spin-flavor symmetry classes.  Such models are computationally
tractable.  Of course, in principle one may use {\em any\/} model
consistent with large $N_c$ scaling to establish this point.  Ideally,
one should consider models that do not suffer from the spurious mode
problem, in order to avoid the issue entirely.  One might consider,
for example, a translationally invariant model of $N_c$ quarks
interacting among themselves.  Such a model obviously suffers the
drawback that it is computationally extremely hard to solve.  However,
for our purposes the only issue of relevance is whether quark spin is
a good quantum number.  If the model has tensor interactions between
quarks at leading order [which is $1/N_c$, leading to $O(N_c^0)$
matrix elements when quark combinatorics are included], then quark
spin is not a good quantum number, and even in the absence of explicit
computation one expects generic mixing of $O(N_c^0)$ between states of
different spin-flavor symmetry.


\begin{thebibliography}{99}
\bibitem{Dwidth}
This includes the $\Delta$; note that it only decays strongly
because the $O(1/N_c)$ $\Delta$-$N$ mass difference is larger
than $m_\pi$: For the case $N_c \!  = \!  3$, the chiral limit
wins out over the large $N_c$ limit.
\end{thebibliography}
\end{document}